# Reaction-Network-Level Discovery of Ammonia Synthesis Catalysts via Ten-Million-Scale Generative Exploration

*Ruili Li[1, 5], Rui Qi[2, 3*], Shuoqi Zhang[1, 5], Qingli Tang[2,3], Qingqing Mao[4], Ritankar Das[4], Beien Zhu[2, 3*], and Yi Gao[2, 3*]*

[1]Key Laboratory of Interfacial Physics and Technology, Shanghai Institute of Applied Physics, Chinese Academy of Sciences, Shanghai 201800, China

[2]Photon Science Research Center for Carbon Dioxide, Shanghai Advanced Research Institute, Chinese Academy of Sciences, Shanghai 201210, China

[3]State Key Laboratory of Low Carbon Catalysis and Carbon Dioxide Utilization, Shanghai Advanced Research Institute, Chinese Academy of Sciences, Shanghai 201210, China

[4]Titan Holdings, Hayward, CA 94541, United States

[5]University of Chinese Academy of Sciences, Beijing 100049, China

*To whom correspondence should be addressed. Email: qir@sari.ac.cn; zhube@sari.ac.cn; gaoyi@sari.ac.cn

ABSTRACT: Catalyst discovery for ammonia synthesis is inherently a reaction-network challenge because catalytic performance is governed not by a single adsorbed intermediate, but by a surface's orchestrated compatibility with multiple distinct intermediates across competing dissociative and associative pathways. However, navigating ultra-large chemical spaces under such multi-intermediate constraints remains a formidable bottleneck for conventional screening workflows. Here, we report a reaction-network-level catalyst discovery framework driven by ten-million-scale generative exploration. By coupling adsorbate-specific generative Transformers with high-throughput machine learning potentials, we systematically map the structure-property landscapes of four critical intermediates (N*, NH*, NNH*, and HNNH*). Scale-dependent overlap analysis shows that the full four-intermediate compatibility space remains strongly under-sampled at conventional $10^5$-$10^6$ generative scales, emerging exclusively under ten-million-scale exploration. By generating approximately 15 million configurations per adsorbate, followed by structural compression and machine-learning-potential predictions, we identified 279 highly potential target materials. This sparse compatibility space successfully recovers traditional Fe- and Ru-based motifs while uncovering previously unexplored catalyst families. Representative DFT calculations validate pathway-dependent mechanisms: Fe-V emerges as a dissociative-pathway lead by significantly lowering the initial $N_2$ dissociation barrier, whereas Al-Pd-Zr efficiently stabilizes associative intermediates as an associative-pathway lead. These findings establish multi-intermediate reaction-network compatibility as a robust criterion for discovering advanced catalysts from multi-million generative chemical spaces.

## Introduction

Catalysis underpins modern chemical manufacturing, yet the discovery of high-performance catalysts remains constrained by the need to navigate vast multidimensional chemical spaces defined by composition, surface structure, and reaction intermediates.[1-3] Traditionally, catalyst development has relied heavily on empirical trial-and-error strategies that are labor-intensive, time-consuming, and inherently limited in chemical scope.[4] In recent years, machine-learning-assisted high-throughput screening has substantially accelerated catalyst evaluation by enabling rapid prediction of adsorption energies and related descriptors across large catalyst libraries.[5-15] However, these approaches remain fundamentally limited by the size and diversity of the candidate databases from which screening begins.

Generative modeling offers a complementary route by expanding catalyst exploration beyond predefined databases toward previously unsampled structures, such as variational autoencoders, generative adversarial networks, and diffusion models.[16-22] In particular, Transformer-based large language models (LLMs) have shown strong potential for materials design because they can directly generate structural representations from textual or sequence-based inputs.[23-30] In heterogeneous catalysis, the generative design framework MAGECS demonstrated large-scale generative exploration for $CO_2$ reduction by producing hundreds of thousands of catalyst structures[31] while pretrained generative Transformers (e.g., CatGPT) enabled adsorbate-conditioned generation of catalyst surfaces. For example, OOH*-conditioned generation for the oxygen reduction reaction produced tens of thousands of candidates and identified high-performance catalysts.[32] More recently, we developed a distributed generative-Transformer framework for ten-million-scale catalyst exploration, using $CH_3$ as a representative single adsorbate for methane-conversion catalyst screening.[33] This work demonstrated that adsorbate-conditioned generative models can substantially expand the accessible catalyst space and enable large-scale material prioritization. Nevertheless, such single-intermediate screening remains insufficient for reactions governed by multiple coupled elementary steps. Industrially relevant catalytic reactions often involve complex reaction networks, in which catalytic performance depends on the balanced stabilization of several distinct intermediates rather than the adsorption energy of one descriptor species alone.

Ammonia synthesis provides a particularly stringent test case for catalyst discovery in complex reaction networks.[34-40] Unlike reactions that can be approximately rationalized by a single adsorption descriptor, ammonia formation involves both dissociative and associative pathways, with distinct elementary steps imposing different energetic and structural requirements on the catalyst surface.[41-45] In the dissociative pathway, $N_2$ first dissociates to surface-bound N*, which is then hydrogenated stepwise through species such as NH* to form $NH_3$. In the associative pathway, hydrogenation begins before N-N bond cleavage, proceeding through intermediates such as NNH* and HNNH* prior to further bond breaking and hydrogenation. As a result, N*, NH*, NNH*, and HNNH* represent not merely four adsorbates, but four chemically distinct constraints that span the major mechanistic branches of the ammonia-synthesis network.[46, 47] A catalyst optimized for a single intermediate often suffers from over-binding or under-binding of others, rendering ammonia synthesis fundamentally a multi-intermediate, multi-pathway compatibility challenge.

Here we address this challenge by formulating catalyst discovery for ammonia synthesis at the reaction-network level rather than at the level of a single adsorbate. Using adsorbate-specific fine-tuning of a pretrained generative Transformer, we construct ultra-large catalyst spaces for N*, NH*, NNH*, and HNNH*, followed by unified structural compression and machine-learning-potential-based energetic evaluation. This strategy enables systematic identification of the sparse materials space that remains compatible across multiple key intermediates under full-network constraints, thereby providing access to catalyst families that would remain hidden in smaller or single-intermediate searches. By explicitly tracking how the accessible space contracts from individual intermediates to coupled multi-intermediate overlap, we demonstrate that adsorbate-resolved exploration at the ten-million scale is a prerequisite to recover chemically meaningful, four-intermediate-compatible materials. Representative DFT evaluation confirms that the recovered overlap space is not merely an adsorption-level compatibility space, but a focused source of pathway-specific ammonia-synthesis catalyst leads. Fe-V emerges as a representative dissociative-pathway candidate with a low $N_2$ activation barrier, whereas Al-Pd-Zr provides a representative associative-pathway lead, capable of stabilizing key hydrogenated $N_2$ intermediates. Additional transition-state calculations on Al-Ti, Au-Ti, and Cu-Zr further reveal multiple material families with pathway-dependent catalytic potential. Broadly, this work provides a practical, reaction-network-guided paradigm to discover and prioritize potential catalysts from previously inaccessible chemical spaces.

**Results and Discussion**

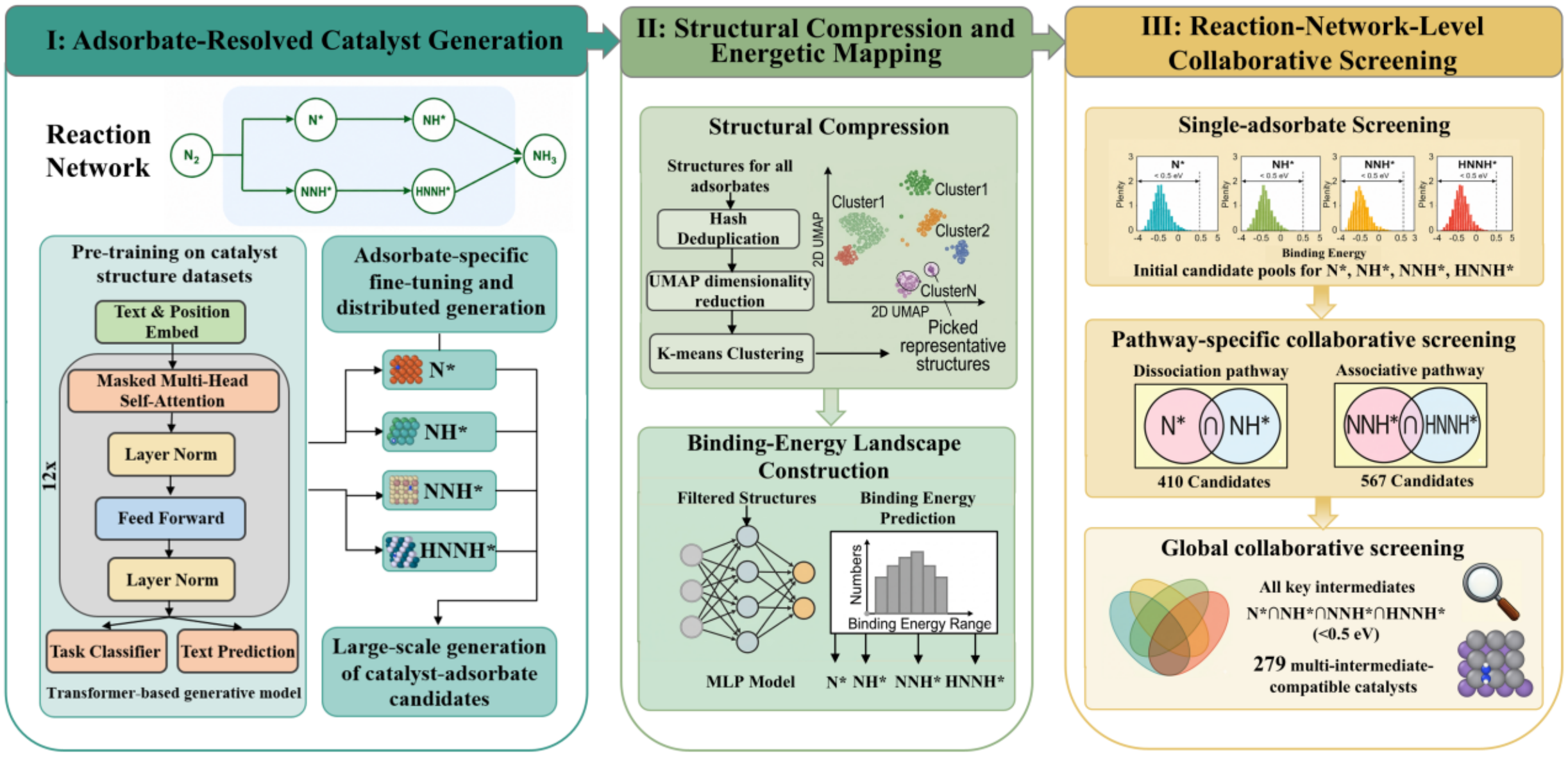

**Figure 1.** Reaction-network-level catalyst discovery for ammonia synthesis.

Figure 1 outlines the reaction-network-centered workflow developed in this work for large-scale exploration of catalyst space and catalyst discovery in complex reaction networks, such as ammonia synthesis. In Step I, a Transformer-based generative model was pretrained on the Open Catalyst 2020 Structure to Energy and Forces dataset (OC20-S2EF 2M),[48] which contains approximately two million catalyst structures with diverse adsorbates and broad compositional coverage. The model consists of 12 layers with a hidden dimension of 512, 8 attention heads, and a maximum sequence length of 1024, with catalyst structures encoded as tokenized sequences of lattice parameters, atomic species, and atomic coordinates. The pretrained model was then underwent four-fold fine-tuning specifically tailored to four critical reaction intermediates (N*, NH*, NNH*, and HNNH*), enabling generation within intermediate-specific chemical spaces while preserving broad structural diversity. To sustain exploration at the ten-million scale, the fine-tuned models were deployed using massively parallel GPU-accelerated generation, producing approximately 15 million candidate structures per adsorbate. In Step II, the generated structural space was compressed through hash-based deduplication, dimensionality reduction, and clustering-assisted selection, thereby removing redundant structures and retaining representative candidates for further evaluation. These selected structures were then assessed using a machine-learning-potential (MLP) EquiformerV2 model to predict binding energies and construct intermediate-specific structure-property landscapes. In Step III, hierarchical collaborative screening was executed to identify promising ammonia-synthesis catalysts by evaluating both pathway-specific and global intermediate coupling. It includes the N*-NH* overlap for the dissociative pathway, the NNH*-HNNH* overlap with the associative pathway, and the final global overlap across all four intermediates. This workflow enables identification of materials that satisfy energetic constraints across multiple key intermediates and thereby defines a chemically meaningful candidate space for subsequent DFT-based prioritization of potential ammonia-synthesis catalysts.

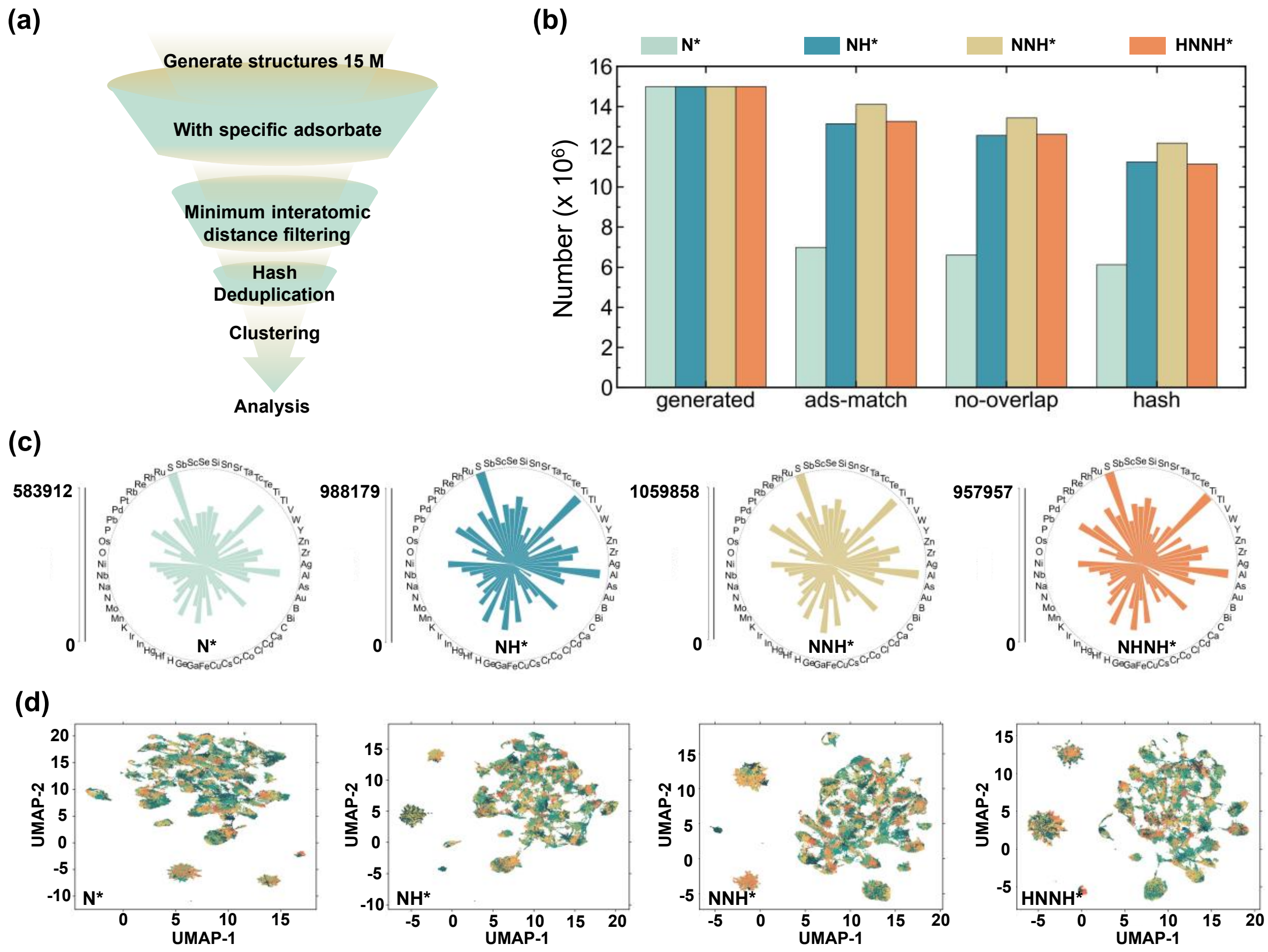

**Figure 2.** Structural compression, elemental distribution, and low-dimensional representation of the generated candidates for the N*, NH*, NNH*, and HNNH* adsorbate systems. (a) Schematic illustration of the structural compression workflow. (b) Retained candidate counts after progressive adsorbate matching, geometric validation, and hash-based deduplication. (c) Elemental distributions in the hash-retained structure sets for the four adsorbate systems, visualized as radial bar plots. (d) UMAP projections of the retained structure spaces.

Based on the pre-trained model, we executed four independent fine-tuning processes using intermediate-specific training sets containing 7,000 N*, 10,000 NH*, 13,000 NNH*, and 13,000 HNNH* structures, respectively (validation across varying dataset sizes is provided in Figure S1). Coupled with distributed generation, the resulting fine-tuned models produce approximately 15 million candidate configurations for each adsorbate-defined system. To manage this multi-million-scale dataset, the raw structures were processed through a hierarchical compression pipeline (Figure 2a). Structures containing the target adsorbate were first retained, followed by geometric filtering to remove unreasonable configurations and subsequent hash-based deduplication and clustering to obtain representative structure sets for further analysis (see Supporting Information Methods, Figure S2 and Table S1). After hash-based deduplication, 6,123,882 N*, 11,243,060 NH*, 12,174,534 NNH*, and 11,136,924 HNNH* structures were retained (Figure 2b). Although the retained numbers differ among the four adsorbate-defined systems, all remain sufficiently large to preserve broad coverage of the accessible structure spaces.

The retained structures were further characterized in terms of composition and structural organization. Radial bar plots show that the generated candidates span a broad compositional space with substantial elemental diversity (Figure 2c, Figure S3). To examine how these structures are organized in local-environment space, we performed UMAP analysis followed by clustering (see Supporting Information Methods). All hash-retained structures were then assigned to these clusters. The resulting UMAP projections (Figure 2d) reveal broad but clearly nonuniform distributions, indicating that the retained structures occupy recurrent regions of structural space rather than forming featureless clouds. Cluster-colored projections further reveal multiple distinct regions within each adsorbate-defined space. Importantly, the distinct organizational topologies observed across the N*, NH*, NNH*, and HNNH* spaces demonstrate that each intermediate imposes unique chemical and geometric constraints on its underlying local coordination environments. These results collectively confirm that our adsorbate-specific generation, coupled with rigorous compression, preserves vast chemical diversity while retaining meaningful structural organization, establishing a solid foundation for downstream MLP-based energetic evaluations and reaction-network-level screening.

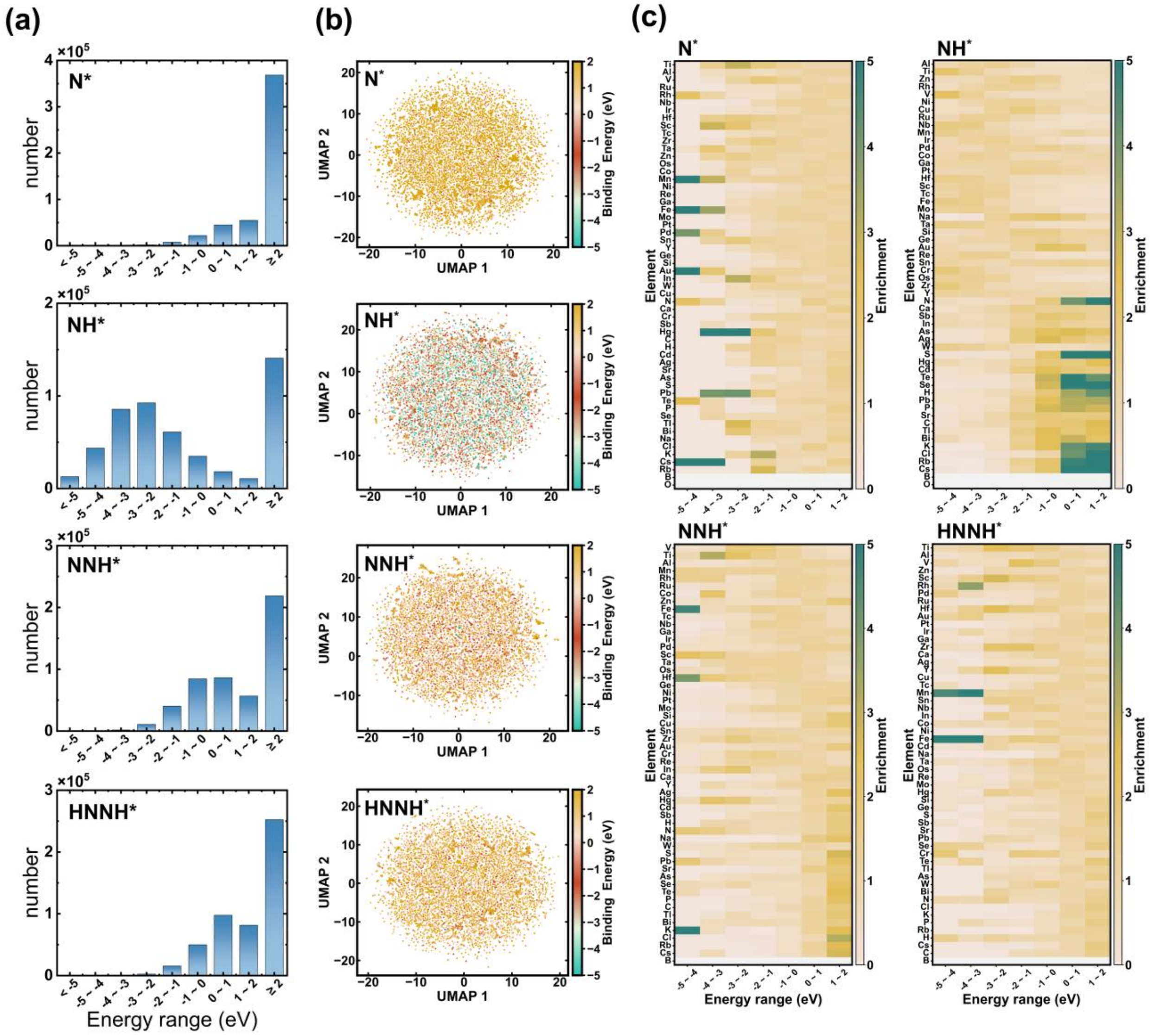

**Figure 3.** Binding-energy landscapes and composition-property relationships of the screened candidates for the N*, NH*, NNH*, and HNNH* adsorbate-defined systems. (a) Corrected binding-energy distributions of the prioritized candidates for adsorbate-defined systems. (b) UMAP projections of the screened catalyst spaces colored by EquiformerV2-predicted binding

energies. (c) Element-wise enrichment across binding-energy intervals. For visualization clarity, enrichment factors exceeding 5 are clipped to 5 in the color scale.

To bridge structural diversity with thermodynamic screening without incurring prohibitive computational costs, we partitioned the hash-retained structures into 200 distinct clusters. We randomly sampled 1% of the structures from each cluster for binding-energy prediction using EquiformerV2 model. Because the reference energies of NH*, NNH*, and HNNH* in OC20-S2EF are defined through linear combinations of N- and H-containing fragments, the raw MLP-predicted values show systematic offsets relative to the adsorption energies directly calculated by DFT in this work. To align all four intermediate landscapes on a unified thermodynamic scale, rigid empirical correction terms of -3.47, -1.64, and -1.28 eV were applied to the inference outputs of NH*, NNH*, and HNNH*, respectively. Guided by these corrected energy profiles, cluster-level scores were assigned and used to guide subsequent selection. Cluster-level scores derived from these predictions were then used to guide subsequent selection, yielding 500,000 retained structures for each adsorbate-defined system while preserving broad coverage of the accessible structural space (see Supporting Information for selection details). Figure 3a illustrates the corrected binding-energy distributions of the retained candidate structures for the N*, NH*, NNH*, and HNNH* systems. While all four adsorbate landscapes occupy relatively compact energy windows, their profiles are distinct; NH* exhibits a noticeably broader energy spread, whereas N*, NNH*, and HNNH* feature more localized distributions. These differences indicate that the corresponding intermediates retain distinct energetic sensitivities to variations in local structure and composition. Consistently, the UMAP projections colored by predicted binding energy (Figure 3b) reveal that configurations with similar thermodynamic profiles naturally segregate into distinct, localized regions rather than dispersing randomly, validating that the generative space preserves coherent structure-property relationships after cluster-based down-sampling.

We further quantified elemental enrichment as a function of binding-energy interval (Figure 3c) to connect composition with predicted adsorption behavior. To quantify composition-property trends, we computed a normalized elemental occurrence ratio for each element $i$ within each binding-energy bin $b$, Here, the enrichment factor was defined as:

$$Enrichment(i, b) = \frac{P(i|b)}{P(i|all)}$$

where $P(i \mid b)$ is the fraction of structures in bin $b$ that contain element $i$, and $P(i \mid all)$ is the corresponding fraction in the full selected dataset. Thus, $Enrichment(i, b)$ measures whether an element is overrepresented or underrepresented in a given energy interval relative to its baseline occurrence in the overall candidate pool. This quantity is analogous to an observed-to-expected or fold-enrichment measure.[49] Values greater than 1 indicate enrichment of an element within that energy interval, whereas values below 1 indicate depletion. The enrichment maps reveal clear nonuniform element-specific trends across the binding-energy range, indicating that composition exerts a measurable influence on the predicted adsorption behavior. Notably, Fe, the cornerstone element of traditional Haber-Bosch catalysis,[50-55] is prominently enriched in the strong-binding (more negative) regions. Conversely, elements such as N, S, and B migrate toward the weaker binding intervals, consistent with their typical localized covalent or non-bonding electronic behaviors on these surface environments. The distinct enrichment fingerprints across the four intermediates emphasize the highly adsorbate-specific nature of composition-property relationships, proving that our framework can transform a massive generative space into an interpretable structure-property landscape suitable for full-network cooperative screening.

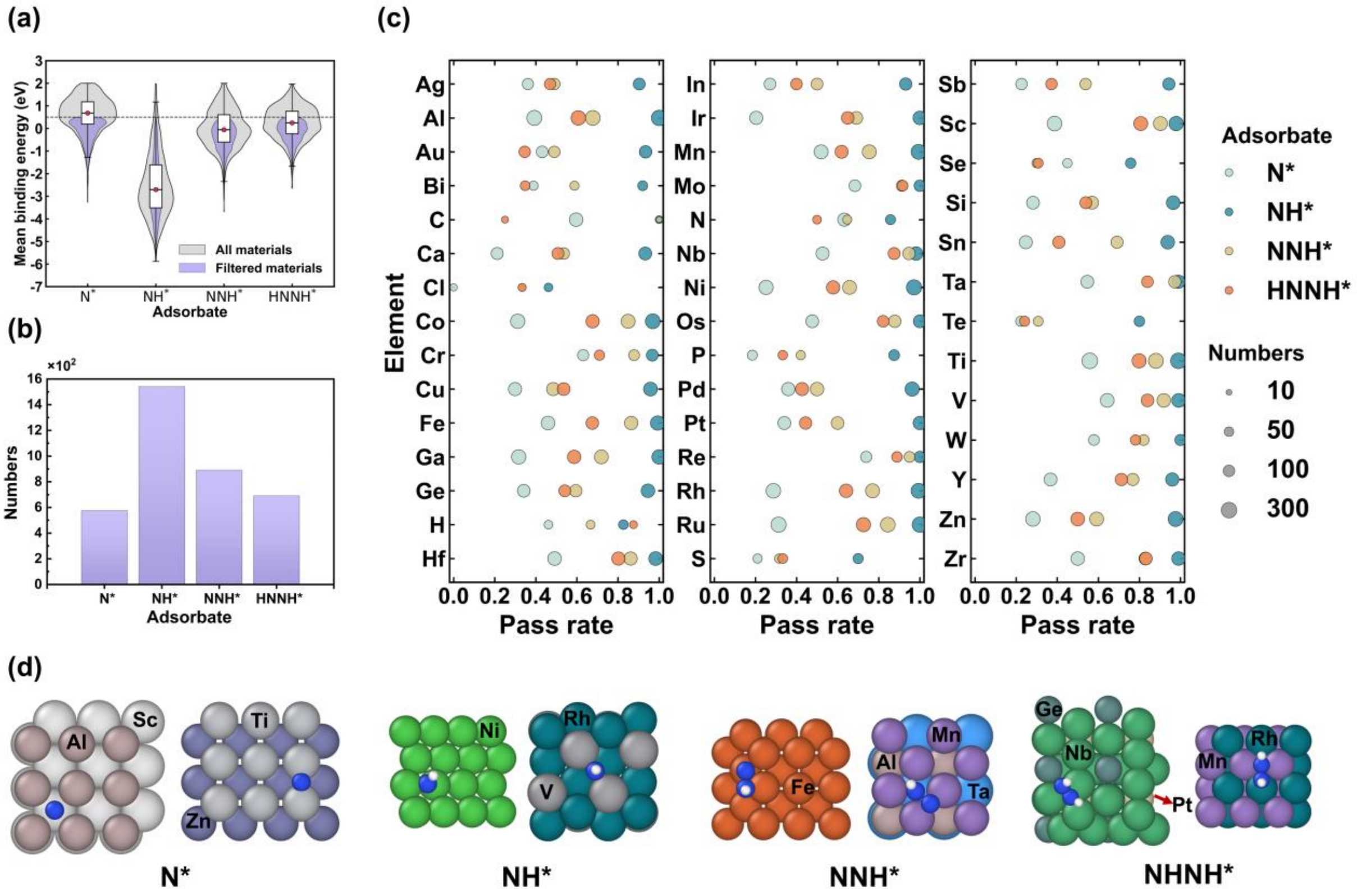

**Figure 4.** Material-level binding-energy statistics and single-adsorbate screening of the generated candidates. (a) Violin plots of material-level mean binding energies across the four adsorbate systems. The horizontal shaded band bounded by dashed lines defines the optimal target energy window (0.5eV) for screening. (b) Total numbers of candidate material families falling within the optimal binding energy window for each adsorbate system. (c) Bubble plots illustrating the element-wise pass rates across the four adsorbate registries. (d) Representative structures from the retained candidate materials across the four adsorbate systems.

To bridge structure-level high-throughput screenings with material-level property evaluations, the compressed structural library was systematically grouped according to elemental composition. Structures containing the same elemental set were assigned to the same material class, irrespective of local stoichiometric variations, thereby establishing a standardized platform for cross-adsorbate comparisons. In this analysis, the 0.5 eV cutoff was used as a permissive screening threshold to retain materials with accessible intermediate stabilization, rather than as a direct activity criterion. This choice allows the workflow to avoid premature elimination of materials that may be relevant under different local configurations or mechanistic branches. As illustrated in Figure 4a, the material-level mean binding energies exhibit clear adsorbate-dependent distributions. NH* shows the strongest overall binding and a broader distribution toward negative energies, whereas N*, NNH*, and HNNH* are mainly distributed around the near-threshold region with distinct profiles. Using the adopted screening criterion of mean binding energy < 0.5 eV, the filtered materials, highlighted by the purple violins, occupy the lower-energy portions of each distribution. The corresponding material counts are summarized in Figure 4b: NH* retains the largest candidate set (1541), followed by NNH* (891) and HNNH* (691), whereas N* gives the smallest retained set (576). These differences indicate that each intermediate imposes distinct energetic selectivity on the generated material space, even before cross-adsorbate coupling is considered. This result further supports the need for multi-intermediate screening rather than relying on a single adsorption descriptor.

The element-wise pass-rate analysis further provides a chemically interpretable view of the screening results (Figure 4c; see Supporting Information for details). Bubble color encodes the pass rate, and bubble size reflects the number of materials containing the corresponding element. The resulting maps show clear adsorbate-dependent compositional preferences, indicating that even single-intermediate screening already imposes element-specific selectivity. Consistent with the larger retained material set in Figure 4b, NH* shows broader element-wise retention, whereas N* exhibits more limited pass-rate coverage. Several transition-metal elements, including Fe, Co, Mn, Mo, Ti, V, and Zr, maintain appreciable pass rates across multiple adsorbates with non-negligible sample support. The presence of Fe-, Co-, Mn-, and Mo-containing candidates is consistent with experimentally reported ammonia-synthesis-related catalysts, suggesting that the screening workflow recovers chemically meaningful motifs rather than random compositional artifacts. [50, 56-60] At the same time, the adsorbate-dependent differences in pass rate show that no single element or single intermediate is sufficient to define the full candidate space. Representative retained structures are shown in Figure 4d, linking the material-level statistics to concrete atomic configurations.

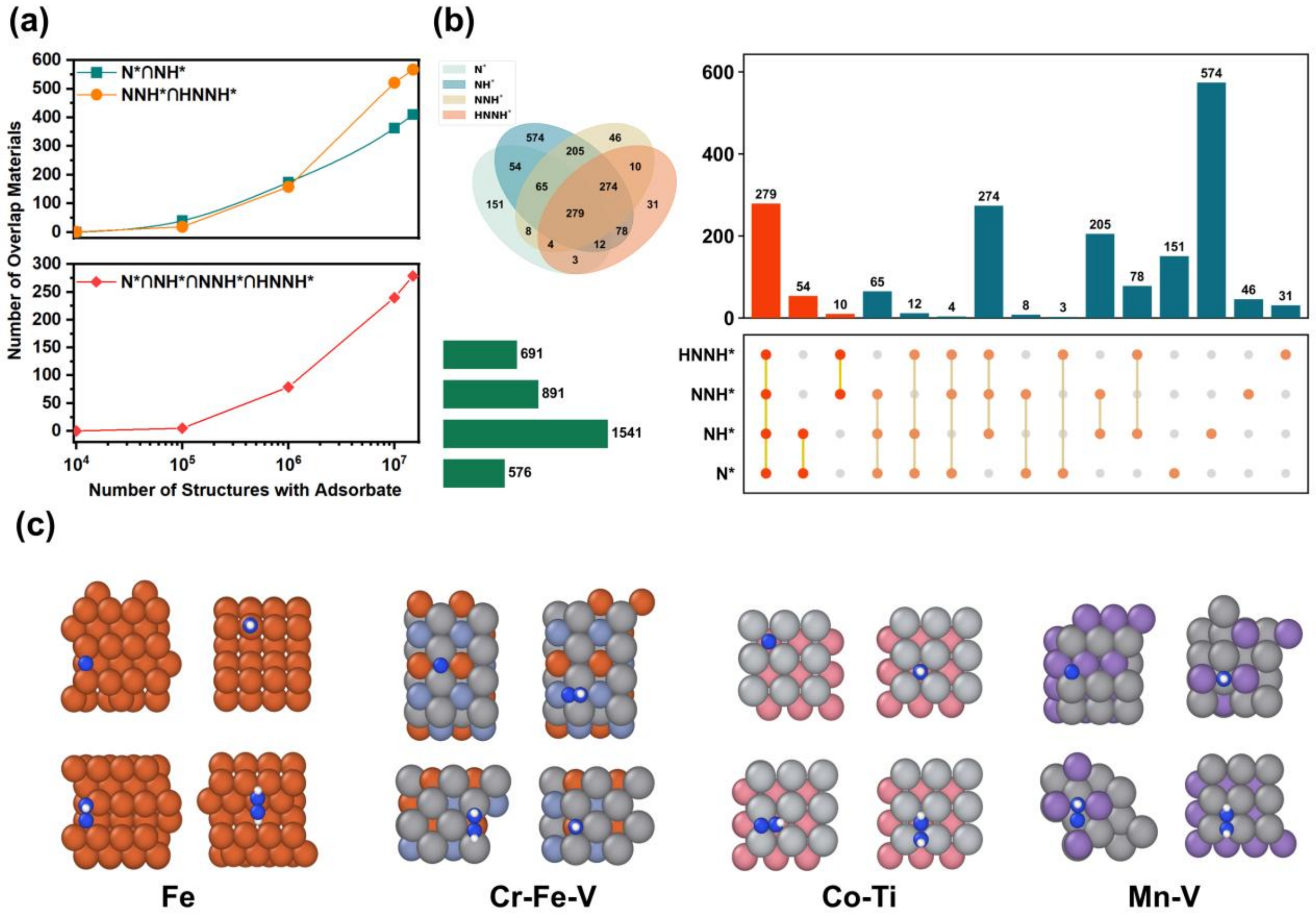

**Figure 5.** Scale-dependent recovery of cross-adsorbate overlap space and representative four-intermediate-compatible candidate materials. (a) Scale-dependent growth of pairwise (N∩NH and NNH*∩HNNH*) and four-intermediate overlap materials. (b) Venn and UpSet analysis of candidate-material overlap across N*, NH*, NNH*, and HNNH* systems. (c) Representative structures of selected four-intermediate-compatible candidate materials.

Having established intermediate-specific material pools, we then examined whether any material families remain compatible across multiple reaction-network constraints. The scale-dependent overlap analysis shows that the compatibility space is extremely sparse. At conventional $10^5$-$10^6$ generation scales, pairwise and four-intermediate overlaps are either absent or severely under-sampled (Figure 5a). Only when the adsorbate-resolved spaces reach the ten-million scale does the full-network overlap become statistically recoverable. This observation provides an important practical message: for reaction networks involving multiple chemically distinct intermediates, insufficient generation scale can lead not only to fewer candidates, but to a qualitatively incomplete view of the compatible catalyst space. The Venn and UpSet analyses in Figure 5b further summarize the overlap relationships among the four adsorbate-defined candidate sets. Because the UpSet bars represent exclusive intersections, the true pairwise overlaps are obtained by summing the corresponding higher-order intersections rather than reading the binary intersections alone. At the pairwise level, 410 materials are retained in the N*∩NH* set and 567 materials in the NNH*∩HNNH* set, corresponding to dissociative- and associative-pathway compatibility, respectively. When all four adsorbate constraints are applied simultaneously, only 279 materials remain in the final overlap space (Figure 5b). Notably, the four-intermediate overlap recovers several experimentally relevant ammonia-synthesis-related motifs. It includes Fe,[51, 54, 55] Ru,[61-63] Fe-Pt,[52] additional Cr-,[64] Ti-,[65] and V-containing compositions,[66] as well as Fe-Ti-[53] and Ru-Ti-[67] based materials. Representative structures shown in Figure 5c further illustrate that the final overlap space retains chemical and structural diversity despite its sparsity. These results demonstrate that ten-million-scale adsorbate-resolved generation is necessary to recover the sparse reaction-network-compatible space, which provides a focused candidate pool for subsequent DFT validation.

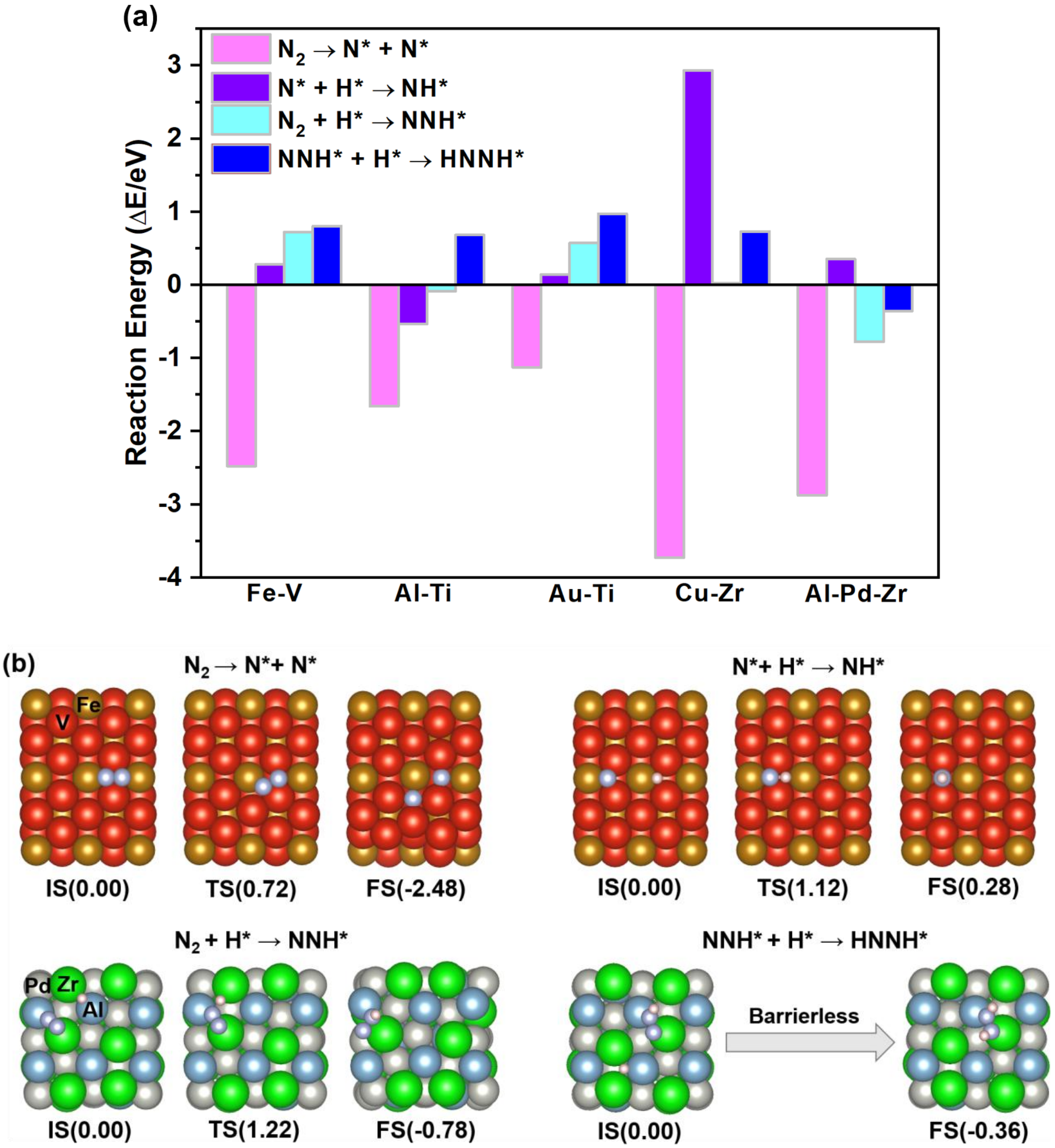

Figure 6. DFT reaction-energy screening and pathway-resolved transition-state validation of representative four-intermediate-overlap material families. (a) Reaction-energy differences of key elementary steps along dissociative and associative ammonia-synthesis pathways. Negative values

indicate exergonic steps, whereas positive values indicate endergonic steps. (b) Optimized initial-state (IS), transition-state (TS), and final-state (FS) structures for representative dissociative steps on Fe-V and associative steps on Al-Pd-Zr. The calculated activation barriers are shown below each elementary step. All reaction-energy differences and activation barriers are given in eV.

To determine whether the derived materials are favorable for elementary steps relevant to ammonia synthesis, we performed DFT reaction-energy and transition-state calculations on representative candidates. Because the overlap screening is performed at the material-composition level, the dissociative pathway and the associative pathway may correspond to different local environments within the same material family. Therefore, the DFT results are interpreted as material-family-level pathway validation, rather than as a strict same-surface microkinetic comparison. Before the DFT validation, we first examined the synthetic accessibility of the selected candidate materials. Literature reports show that these compositions have been synthesized or reported in related alloy/intermetallic forms, supporting their relevance as experimentally accessible material families (Table S2). We then optimized N*, NH*, NNH*, and HNNH* adsorption configurations on selected candidates (Figure S4 and Table S3) and evaluated key reaction-energy differences along dissociative and associative ammonia-synthesis pathways. Figure 6a and Table S4 summarize the reaction-energy differences for representative elementary steps along dissociative and associative pathways. For the dissociative branch, we considered $N_2$ dissociation to N* + N* and subsequent N* + H* → NH* formation. For the associative branch, we examined $N_2$ + H* → NNH* and NNH* + H* → HNNH* formation. The reaction-energy map in Figure 6a reveals distinct pathway-dependent thermodynamic behavior. Fe-V, Al-Ti, Au-Ti, Cu-Zr, and Al-Pd-Zr all show exergonic $N_2$ dissociation, indicating that dissociated N* species can be stabilized on these surfaces. However, the following N* hydrogenation step varies substantially among the candidates. In particular, Cu-Zr shows strongly exergonic $N_2$ cleavage but highly unfavorable NH* formation, demonstrating that $N_2$ activation alone is insufficient to define a viable dissociative pathway. In contrast, Fe-V combines favorable $N_2$ dissociation with moderate N* hydrogenation energy, whereas its first associative step is endergonic by 0.71 eV; thus, Fe-V was assigned as a dissociative-pathway candidate. For Al-Pd-Zr, both $N_2$ dissociation and the formation of NNH* and HNNH* are thermodynamically favorable, indicating that this material family may support both dissociative and associative branches.

Based on these reaction-energy trends, representative elementary steps were selected for transition-state calculations (Figure 6b and Figure S5). Fe-V exhibits a low $N_2$ dissociation barrier of 0.72 eV, followed by a higher N* + H* → NH* barrier of 1.12 eV, indicating that it can promote initial $N_2$ activation through a dissociative route, while subsequent N-H bond formation remains a key step for further optimization. For Al-Pd-Zr, the facile $N_2$ dissociation is followed by a high N* + H* → NH* barrier of 1.96 eV. In contrast, the associative $N_2$* + H* → NNH* step proceeds with a lower barrier of 1.22 eV, followed by a barrierless NNH* + H* → HNNH* step, supporting Al-Pd-Zr as a representative associative-pathway lead. Additional transition-state calculations for Al-Ti, Au-Ti, and Cu-Zr are provided in Figure S5. Al-Ti shows lower barriers along the associative branch than along the dissociative branch, suggesting that this material family is more likely to support an associative pathway. Au-Ti exhibits relatively high barriers for both dissociative and associative elementary steps, indicating that neither pathway is clearly favored without further local-environment optimization. Cu-Zr further highlights the limitation of using $N_2$ activation alone as a descriptor: although $N_2$ dissociation is thermodynamically favorable and kinetically accessible, the subsequent N* + H* → NH* step is highly unfavorable. Overall, these DFT results demonstrate that the four-intermediate overlap space provides a chemically meaningful candidate pool for pathway-resolved kinetic prioritization, where different material families can be assigned to distinct mechanistic tendencies and further optimized at the local-site level.

## Conclusions

In summary, we develop a reaction-network-centered generative framework for catalyst discovery in ammonia synthesis that enables pathway-aware exploration of ultra-large catalyst spaces across multiple key intermediates. By integrating adsorbate-specific generation, high-throughput structural screening, and machine-learning-potential-based energy evaluation, the workflow supports systematic interrogation of approximately 15 million catalyst-adsorbate configurations for each representative intermediate. Scale-dependent analysis shows that this overlap space remains strongly under-sampled at conventional $10^5$-$10^6$ generative scales and begins to emerge under ten-million-scale exploration, yielding 279 four-intermediate-compatible materials. Representative DFT reaction-energy and transition-state calculations further demonstrate that the recovered overlap space can lead to potential ammonia-synthesis catalyst candidates. Fe-V is identified as a representative dissociative-pathway candidate with a low $N_2$ dissociation barrier, whereas Al-Pd-Zr provides a representative associative-pathway candidate capable of supporting key hydrogenated $N_2$ intermediates. Additional calculations on Al-Ti, Au-Ti, and Cu-Zr further confirm that different material families exhibit distinct pathway preferences and step-specific limitations. These results demonstrate that multi-intermediate reaction-network compatibility can effectively enrich large generative chemical spaces toward mechanistically meaningful catalyst candidates, providing a scalable strategy for catalyst-lead prioritization in complex heterogeneous reaction networks.

## ASSOCIATED CONTENT

**Supporting Information**.

The Supporting Information is available free of charge at .

Computational details, structure-generation and screening analyses, representative generated and optimized catalyst structures, literature-based synthetic-feasibility summary, adsorption-energy comparison, and CINEB reaction-pathway energetics can be found in the Supporting Information (PDF).

## AUTHOR INFORMATION

**Corresponding Author**

**Rui Qi** - Photon Science Research Center for Carbon Dioxide, Shanghai Advanced Research Institute, Chinese Academy of Sciences, Shanghai 201210, China; State Key Laboratory of Low Carbon Catalysis and Carbon Dioxide Utilization, Shanghai Advanced Research Institute, Chinese Academy of Sciences, Shanghai 201210, China; qir@sari.ac.cn

**Beien Zhu** - Photon Science Research Center for Carbon Dioxide, Shanghai Advanced Research Institute, Chinese Academy of Sciences, Shanghai 201210, China; State Key Laboratory of Low Carbon Catalysis and Carbon Dioxide Utilization, Shanghai Advanced Research Institute, Chinese Academy of Sciences, Shanghai 201210, China; Email: zhube@sari.ac.cn

**Yi Gao** - Photon Science Research Center for Carbon Dioxide, Shanghai Advanced Research Institute, Chinese Academy of Sciences, Shanghai 201210, China; State Key Laboratory of Low Carbon Catalysis and Carbon Dioxide Utilization, Shanghai Advanced Research Institute, Chinese Academy of Sciences, Shanghai 201210, China; Email: gaoyi@sari.ac.cn

**Authors**

**Ruili Li** - Key Laboratory of Interfacial Physics and Technology, Shanghai Institute of Applied Physics, Chinese Academy of Sciences, Shanghai 201800, China; University of Chinese Academy of Sciences, Beijing 100049, China;

**Shuo1i Zhang** - Key Laboratory of Interfacial Physics and Technology, Shanghai Institute of Applied Physics, Chinese Academy of Sciences, Shanghai 201800, China; University of Chinese Academy of Sciences, Beijing 100049, China;

**Qingli Tang** - Photon Science Research Center for Carbon Dioxide, Shanghai Advanced Research Institute, Chinese Academy of Sciences, Shanghai 201210, China; State Key Laboratory of Low Carbon Catalysis and Carbon Dioxide Utilization, Shanghai Advanced Research Institute, Chinese Academy of Sciences, Shanghai 201210, China;

**Qingqing Mao -** Titan Holdings, Hayward, CA 94541, United States;

**Ritankar Das -** Titan Holdings, Hayward, CA 94541, United States;

**Author Contributions**

R.Q. initiated the project. B.Z. and Y.G. supervised the project. R.L. performed the calculations and data analysis. R.L., Q.M., and R.D. participated in the analysis. R.Q. wrote the original version. B.Z. and Y.G. revised the manuscript. All authors participated in the discussions.

**Notes**

The authors declare no competing interests.

**ACKNOWLEDGMENT**

This work is supported by National Natural Science Foundation of China (92477105, 92577120), Shanghai Municipal Science and Technology Major Project, and Foundation of the Key Laboratory of Low-Carbon Conversion Science & Engineering, Shanghai Advanced Research Institute, Chinese Academy of Sciences (KLLCCSE-202201Z, SARI, CAS). R. Q. thanks for the Innovation Program of Shanghai Advanced Research Institute, CAS (2025CP007). All calculations were performed at National Supercomputing Center in Shanghai.

## REFERENCES

1. Oganov, A. R.; Pickard, C. J.; Zhu, Q.; Needs, R. J., Structure prediction drives materials discovery. *Nature Reviews Materials* **2019,** *4* (5), 331-348.

2. Liu, X.; Liu, B.; Ding, J.; Deng, Y.; Han, X.; Zhong, C.; Hu, W., Building a Library for Catalysts Research Using High-Throughput Approaches. *Adv. Funct. Mater.* **2022,** *32* (1).

3. Ramirez, A.; Lam, E.; Gutierrez, D. P.; Hou, Y.; Tribukait, H.; Roch, L. M.; Copéret, C.; Laveille, P., Accelerated exploration of heterogeneous CO2 hydrogenation catalysts by Bayesian-optimized high-throughput and automated experimentation. *Chem Catalysis* **2024,** *4* (2).

4. Liu, X.; Liang, J.; Wang, Z.; Li, Q.; Deng, Y.; Wang, H., Building Catalyst Exploration Highways by Integrating High-Throughput and Machine Learning Technologies. *Advanced Energy Materials* **2026,** *16* (6).

5. Butler, K. T.; Davies, D. W.; Cartwright, H.; Isayev, O.; Walsh, A., Machine learning for molecular and materials science. *Nature* **2018,** *559* (7715), 547-555.

6. Ahneman, D. T., Predicting reaction performance in C-N cross-coupling using machine learning (vol 360, pg eaat7648, 2018). *Science* **2018,** *360* (6389), 613-613.

7. Reid, J. P.; Sigman, M. S., Holistic prediction of enantioselectivity in asymmetric catalysis. *Nature* **2019,** *571* (7765), 343-+.

8. Chen, Y.; Huang, Y.; Cheng, T.; Goddard, W. A., III, Identifying Active Sites for CO2 Reduction on Dealloyed Gold Surfaces by Combining Machine Learning with Multiscale Simulations. *J. Am. Chem. Soc.* **2019,** *141* (29), 11651-11657.

9. Unke, O. T.; Chmiela, S.; Sauceda, H. E.; Gastegger, M.; Poltaysky, I.; Schuett, K. T.; Tkatchenko, A.; Mueller, K.-R., Machine Learning Force Fields. *Chem. Rev.* **2021,** *121* (16), 10142-10186.

10. Umer, M.; Umer, S.; Zafari, M.; Ha, M.; Anand, R.; Hajibabaei, A.; Abbas, A.; Lee, G.; Kim, K. S., Machine learning assisted high-throughput screening of transition metal single atom based superb hydrogen evolution electrocatalysts. *Journal of Materials Chemistry A* **2022,** *10* (12), 6679-6689.

11. Esterhuizen, J. A.; Goldsmith, B. R.; Linic, S., Interpretable machine learning for knowledge generation in heterogeneous catalysis. *Nat. Catal.* **2022,** *5* (3), 175-184.

12. Bozal-Ginesta, C.; Pablo-Garcia, S.; Choi, C.; Tarancon, A.; Aspuru-Guzik, A., Developing machine learning for heterogeneous catalysis with experimental and computational data. *Nature Reviews Chemistry* **2025,** *9* (9), 601-616.

13. Qi, R.; Zhu, B.; Han, Z.-K.; Gao, Y. High-Throughput Screening of Stable Single-Atom Catalysts in $CO_2$ Reduction Reactions. *ACS Catal.* **2022,** *12*, 8269-8278.

14. Han, Z.-K.; Sarker, D.; Ouyang, R.; Mazheika, A.; Gao, Y.; Levchenko, S. V. Single-Atom Alloy Catalysts Designed by First-Principles Calculations and Artificial Intelligence. *Nat. Commun.* **2021,** *12*, 1833.

15. Zhang, S.; Lu, S.; Zhang, P.; Tian, J.; Shi, L.; Ling, C.; Zhou, Q.; Wang, J., Accelerated Discovery of Single-Atom Catalysts for Nitrogen Fixation via Machine Learning. *Energy and Environmental Material* **2023,** *6* (1).

16. De Breuck, P.-P.; Wang, H.-C.; Rignanese, G.-M.; Botti, S.; Marques, M. A. L., Generative AI for crystal structures: a review. *Npj Computational Materials* **2025,** *11* (1).

17. Kingma, D. P.; Welling, M. Auto-Encoding Variational Bayes 2013, p. arXiv:1312.6114.

18. Rezende, D. J.; Mohamed, S. In *Variational Inference with Normalizing Flows*, 32nd International Conference on Machine Learning, Lille, FRANCE, 2015
Jul 07-09; Lille, FRANCE, 2015; pp 1530-1538.

19. Zhao, L.; Zong, H., AI-Driven Decoding of Material Dynamics: From Machine Learning Potentials and Interpretability to Generative Prediction. *Adv. Mater.* **2025**.

20. Croitoru, F.-A.; Hondru, V.; Ionescu, R. T.; Shah, M., Diffusion Models in Vision: A Survey. *Ieee Transactions on Pattern Analysis and Machine Intelligence* **2023,** *45* (9), 10850-10869.

21. Yan, D.; Smith, A. D.; Chen, C.-C., Structure prediction and materials design with generative neural networks. *Nature Computational Science* **2023,** *3* (7), 572-574.

22. Zeni, C.; Pinsler, R.; Zuegner, D.; Fowler, A.; Horton, M.; Fu, X.; Wang, Z.; Shysheya, A.; Crabbe, J.; Ueda, S.; Sordillo, R.; Sun, L.; Smith, J.; Nguyen, B.; Schulz, H.; Lewis, S.; Huang, C.-W.; Lu, Z.; Zhou, Y.; Yang, H.; Hao, H.; Li, J.; Yang, C.; Li, W.; Tomioka, R.; Xie, T., A generative model for inorganic materials design. *Nature* **2025,** *639* (8055).

23. Boiko, D. A.; Macknight, R.; Kline, B.; Gomes, G., Autonomous chemical research with large language models. *Nature* **2023,** *624* (7992), 570-+.

24. Bran, A. M.; Cox, S.; Schilter, O.; Baldassari, C.; White, A. D.; Schwaller, P., Augmenting large language models with chemistry tools. *Nature Machine Intelligence* **2024,** *6* (5).

25. Dagdelen, J.; Dunn, A.; Lee, S.; Walker, N.; Rosen, A. S.; Ceder, G.; Persson, K. A.; Jain, A., Structured information extraction from scientific text with large language models. *Nat. Commun.* **2024,** *15* (1).

26. Jang, Y.; Kim, J.; Ahn, S. Improving Chemical Understanding of LLMs via SMILES Parsing 2025, p. arXiv:2505.16340.

27. Xie, T.; Wan, Y.; Huang, W.; Zhou, Y.; Liu, Y.; Linghu, Q.; Wang, S.; Kit, C.; Grazian, C.; Zhang, W.; Hoex, B. Large Language Models as Master Key: Unlocking the Secrets of Materials Science with GPT 2023, p. arXiv:2304.02213.

28. Flam-Shepherd, D.; Aspuru-Guzik, A. Language models can generate molecules, materials, and protein binding sites directly in three dimensions as XYZ, CIF, and PDB files 2023, p. arXiv:2305.05708.

29. Jablonka, K. M.; Schwaller, P.; Ortega-Guerrero, A.; Smit, B., Leveraging large language models for predictive chemistry. *Nature Machine Intelligence* **2024,** *6* (2), 161-169.

30. Fu, N.; Wei, L.; Song, Y.; Li, Q.; Xin, R.; Omee, S. S.; Dong, R.; Siriwardane, E. M. D.; Hu, J., Material transformers: deep learning language models for generative materials design. *Machine Learning: Science and Technology* **2023,** *4* (1).

31. Song, Z.; Fan, L.; Lu, S.; Ling, C.; Zhou, Q.; Wang, J., Inverse design of promising electrocatalysts for CO<sub>2</sub> reduction via generative models and bird swarm algorithm. *Nat. Commun.* **2025,** *16* (1).

32. Mok, D. H.; Back, S., Generative pretrained transformer for heterogeneous catalysts. *J. Am. Chem. Soc.* **2024,** *146* (49), 33712-33722.

33. Li, R.; Zhang, S.; Tang, Q.; Mao, Q.; Das, R.; Qi, R.; Zhu, B.; Gao, Y. Generative intelligence explores the chemical space of ten million catalysts. *Chem. Sci.* **2026,** Advance Article.

34. Suryanto, B. H. R.; Matuszek, K.; Choi, J.; Hodgetts, R. Y.; Du, H.-L.; Bakker, J. M.; Kang, C. S. M.; Cherepanov, P. V.; Simonov, A. N.; MacFarlane, D. R., Nitrogen reduction to ammonia at high efficiency and rates based on a phosphonium proton shuttle. *Science* **2021,** *372* (6547), 1187-+.

35. Iriawan, H.; Andersen, S. Z.; Zhang, X.; Comer, B. M.; Barrio, J.; Chen, P.; Medford, A. J.; Stephens, I. E. L.; Chorkendorff, I.; Shao-Horn, Y., Methods for nitrogen activation by reduction and oxidation. *Nature Reviews Methods Primers* **2021,** *1* (1).

36. Chen, J. G.; Crooks, R. M.; Seefeldt, L. C.; Bren, K. L.; Bullock, R. M.; Darensbourg, M. Y.; Holland, P. L.; Hoffman, B.; Janik, M. J.; Jones, A. K.; Kanatzidis, M. G.; King, P.; Lancaster, K. M.; Lymar, S. V.; Pfromm, P.; Schneider, W. F.; Schrock, R. R., Beyond fossil fuel-driven nitrogen transformations. *Science* **2018,** *360* (6391).

37. Fu, X.; Zhang, J.; Kang, Y., Recent advances and challenges of electrochemical ammonia synthesis. *Chem Catalysis* **2022,** *2* (10), 2590-2613.

38. Fu, X.; Pedersen, J. B.; Zhou, Y.; Saccoccio, M.; Li, S.; Salinas, R.; Li, K.; Andersen, S. Z.; Xu, A.; Deissler, N. H.; Mygind, J. B. V.; Wei, C.; Kibsgaard, J.; Vesborg, P. C. K.; Norskov, J. K.; Chorkendorff, I., Continuous-flow electrosynthesis of ammonia by nitrogen reduction and hydrogen oxidation. *Science* **2023,** *379* (6633), 707-712.

39. Qing, G.; Ghazfar, R.; Jackowski, S. T.; Habibzadeh, F.; Ashtiani, M. M.; Chen, C.-P.; Smith, M. R., III; Hamann, T. W., Recent Advances and Challenges of Electrocatalytic N<sub>2</sub> Reduction to Ammonia. *Chem. Rev.* **2020,** *120* (12), 5437-5516.

40. Fu, X.; Xu, A.; Pedersen, J. B.; Li, S.; Sazinas, R.; Zhou, Y.; Andersen, S. Z.; Saccoccio, M.; Deissler, N. H.; Mygind, J. B. V.; Kibsgaard, J.; Vesborg, P. C. K.; Norskov, J. K.; Chorkendorff, I., Phenol as proton shuttle and buffer for lithium-mediated ammonia electrosynthesis. *Nat. Commun.* **2024,** *15* (1).

41. Bazhenova, T. A.; Shilov, A. E., NITROGEN-FIXATION IN SOLUTION. *Coord. Chem. Rev.* **1995,** *144*, 69-145.

42. Ertl, G., Reactions at surfaces: From atoms to complexity (Nobel lecture). *Angewandte Chemie-International Edition* **2008,** *47* (19), 3524-3535.

43. McEnaney, J. M.; Singh, A. R.; Schwalbe, J. A.; Kibsgaard, J.; Lin, J. C.; Cargnello, M.; Jaramillo, T. F.; Norskov, J. K., Ammonia synthesis from N2 and H2O using a lithium cycling electrification strategy at atmospheric pressure. *Energy Environ. Sci.* **2017,** *10* (7), 1621-1630.

44. Wang, Q.; Pan, J.; Guo, J.; Hansen, H. A.; Xie, H.; Jiang, L.; Hua, L.; Li, H.; Guan, Y.; Wang, P.; Gao, W.; Liu, L.; Cao, H.; Xiong, Z.; Vegge, T.; Chen, P., Ternary ruthenium complex hydrides for ammonia synthesis via the associative mechanism. *Nat. Catal.* **2021,** *4* (11), 959-+.

45. Talukdar, B.; Kuo, T.-C.; Sneed, B. T.; Lyu, L.-M.; Lin, H.-M.; Chuang, Y.-C.; Cheng, M.-J.; Kuo, C.-H., Enhancement of NH3 Production in Electrochemical N<sub>2</sub> Reduction by the Cu-Rich Inner Surfaces of Beveled CuAu Nanoboxes. *Acs Applied Materials & Interfaces* **2021,** *13* (44), 51839-51848.

46. MacFarlane, D. R.; Cherepanov, P. V.; Choi, J.; Suryanto, B. H. R.; Hodgetts, R. Y.; Bakker, J. M.; Vallana, F. M. F.; Simonov, A. N., A Roadmap to the Ammonia Economy. *Joule* **2020,** *4* (6), 1186-1205.

47. Ruan, Y.; He, Z.-H.; Liu, Z.-T.; Wang, W.; Hao, L.; Xu, L.; Robertson, A. W.; Sun, Z., Emerging two-dimensional materials for the electrocatalytic nitrogen reduction reaction to yield ammonia. *Journal of Materials Chemistry A* **2023,** *11* (42), 22590-22607.

48. Chanussot, L.; Das, A.; Goyal, S.; Lavril, T.; Shuaibi, M.; Riviere, M.; Tran, K.; Heras-Domingo, J.; Ho, C.; Hu, W., Open catalyst 2020 (OC20) dataset and community challenges. *Acs Catalysis* **2021,** *11* (10), 6059-6072.

49. Faver, J. C.; Riehle, K.; Lancia, D. R., Jr.; Milbank, J. B. J.; Kollmann, C. S.; Simmons, N.; Yu, Z.; Matzuk, M. M., Quantitative Comparison of Enrichment from DNA-Encoded Chemical Library Selections. *ACS Comb Sci* **2019,** *21* (2), 75-82.

50. Hattori, M.; Okuyama, N.; Kurosawa, H.; Hara, M., Low-Temperature Ammonia Synthesis on Iron Catalyst with an Electron Donor. *J. Am. Chem. Soc.* **2023,** *145* (14), 7888-7897.

51. Nielsen, A.; Kjaer, J.; Hansen, B., RATE EQUATION AND MECHANISM OF AMMONIA SYNTHESIS AT INDUSTRIAL CONDITIONS. *J. Catal.* **1964,** *3* (1), 68-79.

52. Ahsan, A.; Hussain, M.; Rashad, M. A.; Akhter, P.; Jamil, F.; Cho, K.; Park, Y.-K., Critical review on electrocatalytic reduction of nitrogen and nitrate to ammonia. *Journal of Industrial and Engineering Chemistry* **2025,** *149*, 313-336.

53. Mao, C.; Li, H.; Gu, H.; Wang, J.; Zou, Y.; Qi, G.; Xu, J.; Deng, F.; Shen, W.; Li, J.; Liu, S.; Zhao, J.; Zhang, L., Beyond the Thermal Equilibrium Limit of Ammonia Synthesis with Dual Temperature Zone Catalyst Powered by Solar Light. *Chem* **2019,** *5* (10), 2702-2717.

54. Neurock, M., The microkinetics of heterogeneous catalysis. By J. A. Dumesic, D. F. Rudd, L. M. Aparicio, J. E. Rekoske, and A. A. Treviño, ACS Professional Reference Book, American Chemical Society, Washington, DC, 1993, 315 pp. *AIChE Journal* **1994,** *40* (6), 1085-7.

55. Kim, J.-H.; Dai, T.-Y.; Yang, M.; Seo, J.-M.; Lee, J. S.; Kweon, D. H.; Lang, X.-Y.; Ihm, K.; Shin, T. J.; Han, G.-F.; Jiang, Q.; Baek, J.-B., Achieving volatile potassium promoted ammonia synthesis via mechanochemistry. *Nat. Commun.* **2023,** *14* (1).

56. Gao, W.; Wang, P.; Guo, J.; Chang, F.; He, T.; Wang, Q.; Wu, G.; Chen, P., Barium Hydride-Mediated Nitrogen Transfer and Hydrogenation for Ammonia Synthesis: A Case Study of Cobalt. *Acs Catalysis* **2017,** *7* (5), 3654-3661.

57. Kojima, R.; Aika, K., Cobalt molybdenum bimetallic nitride catalysts for ammonia synthesis Part 2. Kinetic study. *Applied Catalysis a-General* **2001,** *218* (1-2), 121-128.

58. Qu, W.; Roy, P.; Wang, C.; Ma, L.; Bu, F.; Zhang, X.; He, Z.; Tsapatsis, M.; Bukowski, B. C.; Wang, C., Earth-Abundant Manganese Nitride Catalysts for Mild-Condition Ammonia Synthesis. *Acs Catalysis* **2025,** *15* (6), 4817-4823.

59. Wang, J.; Liu, L.; Li, R.; Wang, S.; Ju, X.; He, T.; Guo, J.; Chen, P., Highly active manganese nitride-europium nitride catalyst for ammonia synthesis. *Iscience* **2024,** *27* (9).

60. Volpe, L.; Boudart, M., AMMONIA-SYNTHESIS ON MOLYBDENUM NITRIDE. *J. Phys. Chem.* **1986,** *90* (20), 4874-4877.

61. Kitano, M.; Inoue, Y.; Yamazaki, Y.; Hayashi, F.; Kanbara, S.; Matsuishi, S.; Yokoyama, T.; Kim, S.-W.; Hara, M.; Hosono, H., Ammonia synthesis using a stable electride as an electron donor and reversible hydrogen store. *Nat. Chem.* **2012,** *4* (11), 934-940.

62. Aika, K., Activation of nitrogen by alkali metal promoted transition metal I. Ammonia synthesis over ruthenium promoted by alkali metal. *J. Catal.* **1972,** *27* (3), 424-431.

63. Hattori, M.; Iijima, S.; Nakao, T.; Hosono, H.; Hara, M., Solid solution for catalytic ammonia synthesis from nitrogen and hydrogen gases at 50°C. *Nat. Commun.* **2020,** *11* (1).

64. Guan, Y.; Zhang, W.; Wang, Q.; Weidenthaler, C.; Wu, A.; Gao, W.; Pei, Q.; Yan, H.; Cui, J.; Wu, H.; Feng, S.; Wang, R.; Cao, H.; Ju, X.; Liu, L.; He, T.; Guo, J.; Chen, P., Barium chromium nitride-hydride for ammonia synthesis. *Chem Catalysis* **2021,** *1* (5), 1042-1054.

65. Tricker, A. W.; Hebisch, K. L.; Buchmann, M.; Liu, Y.-H.; Rose, M.; Stavitski, E.; Medford, A. J.; Hatzell, M. C.; Sievers, C., Mechanocatalytic Ammonia Synthesis over TiN in Transient Microenvironments. *ACS Energy Lett.* **2020,** *5* (11), 3362-3367.

66. Cao, Y.; Saito, A.; Kobayashi, Y.; Ubukata, H.; Tang, Y.; Kageyama, H., Vanadium Hydride as an Ammonia Synthesis Catalyst. *Chemcatchem* **2021,** *13* (1), 191-195.

67. Li, C.; Zhang, Z.; Zheng, Y.; Fang, B.; Ni, J.; Lin, J.; Lin, B.; Wang, X.; Jiang, L., Titanium modified Ru/$CeO_2$ catalysts for ammonia synthesis. *Chem. Eng. Sci.* **2022,** *251*.

# Table of the content

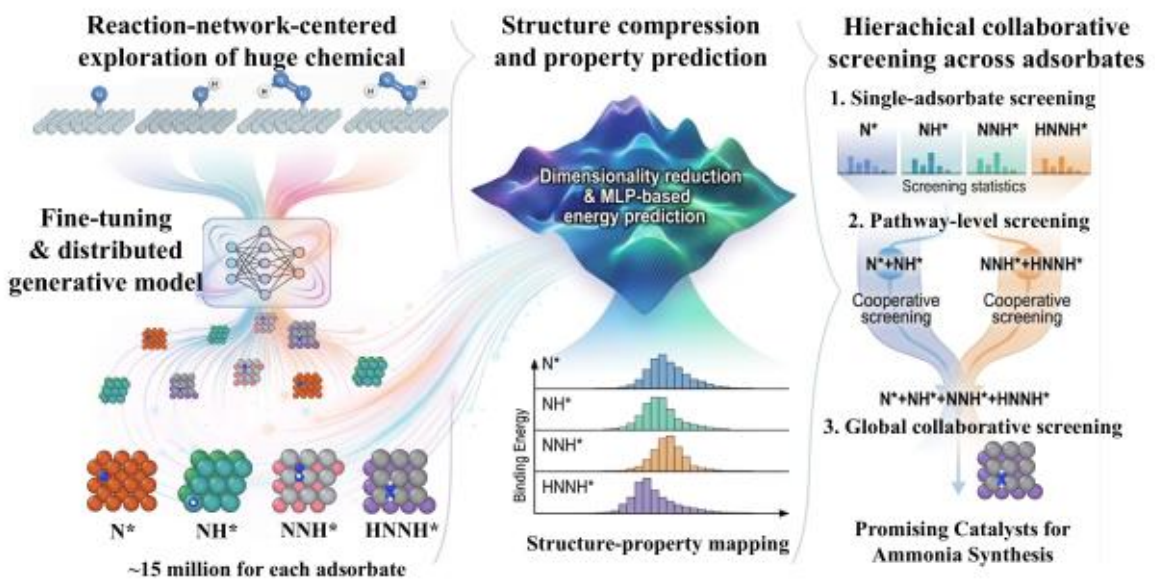

Reaction-Network-Guided Exploration of Ultra-large Chemical Spaces for Ammonia Synthesis